\documentclass[twocolumn,prx,showpacs,superscriptaddress,amssymb,amsmath,amsmath,floatfix]{revtex4-2}
\usepackage{float}
\usepackage{graphicx}
\usepackage{epstopdf}
\usepackage{hyperref}
\usepackage{diagbox}
\usepackage{float}
\usepackage{gensymb}
\usepackage{svg}
\usepackage{times}
\usepackage[normalem]{ulem}
\hypersetup{
   pdfpagemode=None, 
   pdfstartpage=1,
   pdfmenubar=true,
   pdftoolbar=true,
   colorlinks = true,
   linkcolor=blue,
   citecolor=blue,
   urlcolor=blue,
   bookmarksopen=false}

\newcommand{\be}{\begin{equation}}
\newcommand{\ee}{\end{equation}}

\begin{document}{}
\title{Interactions and cold collisions of AlF in the ground and excited electronic states with He}

\author{Sangami Ganesan-Santhi} 
\affiliation{Faculty of Physics, University of Warsaw, Pasteura 5, 02-093 Warsaw, Poland}
\author{Matthew D. Frye}
\affiliation{Faculty of Physics, University of Warsaw, Pasteura 5, 02-093 Warsaw, Poland}
\author{Marcin Gronowski}
\affiliation{Faculty of Physics, University of Warsaw, Pasteura 5, 02-093 Warsaw, Poland}
\author{Micha{\l} Tomza}
\email{michal.tomza@fuw.edu.pl}
\affiliation{Faculty of Physics, University of Warsaw, Pasteura 5, 02-093 Warsaw, Poland}

\date{\today}

\begin{abstract}
Aluminium monofluoride (AlF) is a promising candidate for laser cooling and the production of dense ultracold molecular gases, thanks to its relatively high chemical stability and diagonal Frank-Condon factors. In this study, we examine the interactions and collisions of AlF in its $X^1\Sigma^+$, $a^3\Pi$, and $A^{1}\Pi$ electronic states with ground-state He using state-of-the-art \textit{ab initio} quantum chemistry techniques. We construct accurate potential energy surfaces (PESs) employing either the explicitly correlated coupled-cluster CCSD(T)-F12 method augmented by the CCSDT correction or the multireference configuration-interaction method for higher-excited electronic states. Subsequently, we employ these PESs in coupled-channel calculations to determine the scattering cross-sections for AlF+He collisions and bound states of the complex. We estimate the uncertainty of the calculated PESs and apply it to assess the uncertainty of the scattering results. We find a relatively low sensitivity of the cross-sections to the variation of the PESs, but the positions of shape resonances remain uncertain. The present results are relevant for further improvements and optimizations of buffer-gas cooling of AlF molecules.
\end{abstract}

\maketitle

\section{Introduction}

Ultracold systems are an excellent platform for various experiments ranging from controlled chemical reactions~\cite{BohnScience17} and precision measurements~\cite{SafronovaRMP18} to quantum simulation of many-body physics~\cite{GrossScience17} and quantum information processing~\cite{Carr_2009}. Research into cold and ultracold molecules experienced a significant surge following the first successful production of an ultracold gas of molecules in its absolute ground state~\citep{NiKK2008}. The complex architecture of energy states within molecules, which comprise electronic, vibrational, rotational, fine, and hyperfine degrees of freedom~\cite{Krems2018}, provides opportunities for precise control and manipulation of both internal and translational molecular motion. This capability facilitates comprehensive explorations of chemical interactions, spectroscopy, and fundamental physics~\cite{Carr_2009}.

Ultracold molecules can be produced by two broad strategies. The first involves photo- or magneto-association of pre-cooled atoms at ultralow temperatures~\cite{photoasso2012,kohler2006}, usually followed by optical transfer to a deeply bound state~\cite{VitanovRMP17}. The alternative strategy involves direct cooling of molecules themselves from higher temperatures. Buffer gas cooling~\cite {HutzlerCR12} is a direct cooling method in which the translational and internal energies are dissipated through collisions with cold, inert gas atoms, usually helium. This typically serves as one of the initial cooling steps since it is generally effective at temperatures above a few hundred millikelvin~\cite{Wesley2009} but can produce high densities in preparation for a further cooling stage to ultracold temperatures. In general, buffer-cell cooling allows the thermalization of both translational and internal degrees of freedom; however, the balance between them is determined by the ratio of elastic to inelastic collisions. Laser cooling~\cite{TarbuttCP18} is another direct cooling method and is frequently used as a second-stage cooling method for molecules, paired with buffer gas cooling. In this process, the directional absorption of photons results in slowing down molecules. This technique requires nearly closed optical cycling and is ubiquitous in atom cooling, but it has also been extended to a particular class of molecules with highly diagonal Frank-Condon factors~\cite{Shuman2010,Tarbutt2018}.

The aluminium fluoride (AlF) molecule is an ideal candidate for cooling to ultracold temperatures \cite{Hofsass_2021,Karra2022}. This is primarily due to its highly diagonal Frank-Condon factors, which are perfect for laser cooling \cite{TruppePRA19,WellsPCCP11}. However, another advantage of AlF is its chemical stability due to its relatively high binding energy (about 7~eV). As the binding energy of simple monofluorides grows, the efficiency of the reaction of the fluorinating reagent with hot ablated metal also increases, as was demonstrated by theoretical computations for AlF and CaF \cite{Liu2022}. A comparison of buffer-gas-cooled beams of four monofluorides showed that the one containing AlF was an order of magnitude brighter than others due to this increased stability relative to MgF, CaF, and YbF~\cite{Sidney2022}. New accurate spectroscopic studies of AlF in molecular beams have been recently reported~\cite{TruppePRA19,DOPPELBAUERMP2020,WalterJCP22}, and the first magneto-optical traps with AlF are under construction~\cite{Hofsass_2021}. Once cooled, its relatively large dipole moment of 1.5 D makes it a promising candidate for studying numerous types of dipolar physics.

Before AlF became an object of research at ultralow temperatures, the discovery of AlF in a protoplanetary circumstellar envelope sparked astronomers' interest in this molecule~\cite{Highberger2001}. For this reason, the first potential energy surface of the ground-state AlF+He complex was reported~\cite{GOTOUMAPSS2011}. This potential exhibited one minimum only with the well depth of about 24~cm$^{-1}$. Newer studies~\cite{Karra2022} suggested that this potential is shallower by about 2~cm$^{-1}$.

In this work, we aim to investigate the prospects for buffer-gas cooling of AlF molecules. As a theoretical foundation, we study in detail the interactions between the AlF molecule in its three lowest electronic states ($X^{1}\Sigma^{+}$, $a^{3}\Pi$, and $A^{1}\Pi$) and ground-state He using accurate \textit{ab initio} quantum chemistry methods. We compute two-dimensional potential energy surfaces for the five lowest electronic states of the AlF+He complex. We evaluate the accuracy of our calculations by analyzing the convergence with the wave function quality and basis set size and the impact of the relativistic effects. Next, we employ the electronic stricture data in coupled-channel scattering calculations of elastic and inelastic scattering cross sections. We find the high ratio of elastic to inelastic collisions, suggesting that buffer gas cooling of AlF molecules can be efficient.  

The plan of the paper is as follows. The theory behind the construction of the potential energy surface and a brief account of collision theory are discussed in Section~\ref{sec:methods}. Potential energy surfaces and scattering results are presented in Section~\ref{sec:col}. We conclude our work in Section~\ref{sec:summary}.

\section{Methods}
\label{sec:methods}

The $X^{1}\Sigma^{+}$, $a^{3}\Pi$, and $A^{1}\Pi$ electronic states of AlF upon interaction with ground-state He correspond to the following electronic states of the interacting complex: $X^{1}A'$, $a^{3}A''$, $b^{3}A'$, $A^{1}A'$, and $B^{1}A''$, under the $C_\textrm{s}$ point group, 
as collected in Table~\ref{tab:1}. We use advanced \textit{ab initio} quantum-chemical methods to describe all the mentioned electronic states of AlF+He. Our calculations are based on the Born-Oppenheimer approximation, in which a separate potential energy surface is defined for each electronic state.

\begin{table}[t]
\caption{The relation and symmetry of the electronic states of the AlF+He complex to the electronic states of interacting AlF and He.}
\label{tab:1} 
\begin{ruledtabular}
\begin{tabular}{ccc}
AlF & He & AlF+He \\
\hline
$X^{1}\Sigma^{+}$ & $^1S$ & $X^{1}A'$ \\
$a^{3}\Pi$        & $^1S$& $b^{3}A'$, $a^{3}A''$\\
$A^{1}\Pi$        & $^1S$   & $A^{1}A'$, $B^{1}A''$ \\
\end{tabular}
\end{ruledtabular}
\end{table}

\subsection{\textit{Ab initio} electronic structure methods}
\label{sec:ab}

\begin{figure}[t]
\includegraphics[width=0.7\columnwidth]{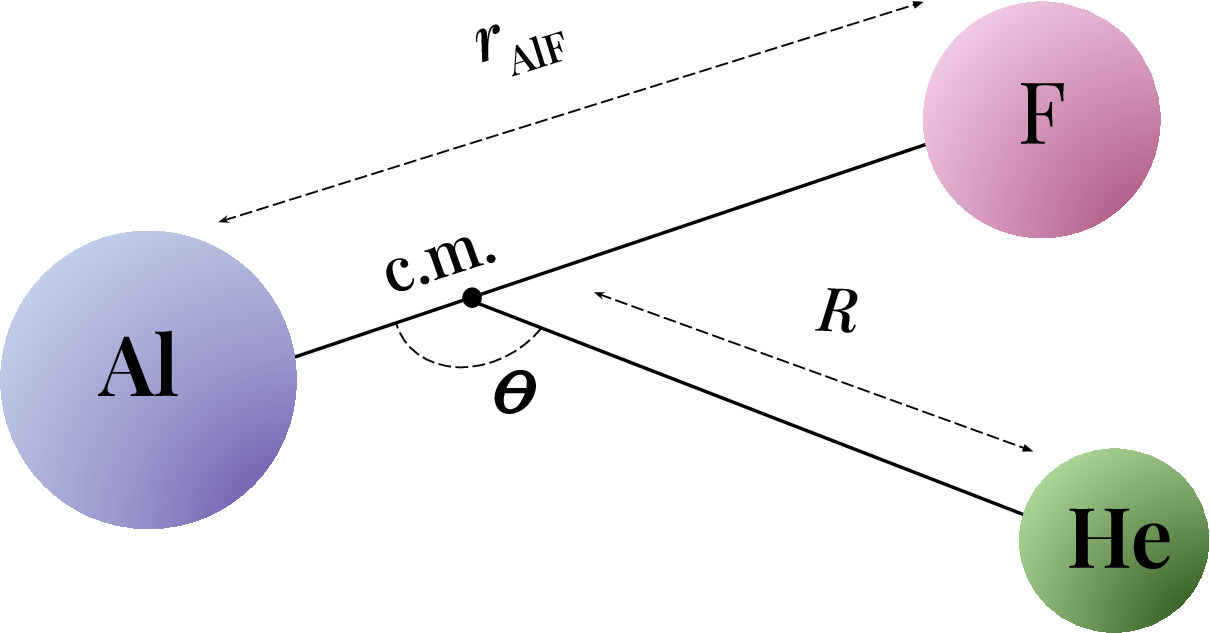}
\caption{The Jacobi coordinates for the AlF+He system.}
\label{fig:1}
\end{figure}

The AlF molecule is considered as a rigid rotor with a fixed bond length $r_\textrm{AlF}$. The Jacobi coordinates $R$ and $\theta$ are used to describe the orientation of the molecule and the atom. $R$ is the distance between the helium atom and the center of mass (c.m.) of the molecule and $\theta$ is the angle between the molecular axis and vector from c.m.~to He ($\theta = 0^{\circ}$ and $\theta = 180^{\circ}$ correspond to the linear HeAlF and AlFHe arrangements, respectively). The coordinates are presented in Fig.~\ref{fig:1}. We use the vibrationally averaged value of the bond length $r_\textrm{AlF}$ for the corresponding electronic states \cite{Jeziorska2000, Faure2016} (3.136$\,$bohr for $X^{1}\Sigma^{+}$, 3.124$\,$bohr for $a^{3}\Pi$, and 3.126$\,$bohr for $A^{1}\Pi$). The interaction potential, $V_\textrm{int}$, depends on both $R$ and $\theta$. We obtain $V_\textrm{int}(R,\theta)$ by using the supermolecular method,
\begin{equation}
V_\textrm{int}(R,\theta) = E_\textrm{AlF+He}(R,\theta)-E_\textrm{AlF}(R,\theta)-E_\textrm{He}(R,\theta),
\end{equation}
where $E_\textrm{AlF+He}$ is the total energy of the complex, while $E_\textrm{AlF}$ and $E_\textrm{He}$ are the energies of the monomers. The basis set superposition error is corrected using the counterpoise correction \cite{Boys1970}, where the monomer energies are also calculated in the same basis set as that of the whole complex. 

The explicitly correlated coupled cluster method \cite{ccsd-f12} restricted to a single, double, and non-iterative triple excitations (CCSD(T)-F12b) \cite{BOKHANPCCP08}, is used to calculate the potential energy surfaces for the $X^{1}A'$, $a^{3}A''$, and $b^{3}A'$ electronic states of the AlF + He complex. In CCSD(T)-F12b computations, we use aug-cc-pV6Z \cite{mourik2000a} as the orbital basis set, aug-cc-pV6Z-RIFIT as the density fitting and resolution of the identity basis set, and aug-cc-pV5Z-JKFIT as the density fitting basis set for the exchange and Fock operators.
We obtain aug-cc-pV5Z-JKFIT by augmenting cc-pV5Z-JKFIT~\cite{weigend2002}. cc-pV5Z-JKFIT for He is based on the unpublished work~\cite{ccRepo}. aug-cc-pV6Z-RIFIT is based on unpublished work available in the Basis Set Exchange repository~\cite{pritchard2019a,feller1996a,schuchardt2007a}. 
We improve the description of the electronic correlation in the $X^{1}A'$, $a^{3}A''$, and $b^{3}A'$ states by including the full triple correction:
\begin{equation}
    \delta V_\textrm{int}^\textrm{CCSDT} = V_\textrm{int}^\textrm{CCSDT}-V_\textrm{int}^\textrm{CCSD(T)}\,,
\end{equation}
where $V_\textrm{int}^\textrm{CCSDT}$ and $V_\textrm{int}^\textrm{CCSD(T)}$ are the interaction energies calculated using the the CCSDT and CCSD(T) methods, respectively, with the aug-cc-pVTZ basis set \cite{Woon1994,Woon1993,Kendall1992}.

We describe $A^{1}A'$ and $B^{1}A''$ states of the complex with the internally contracted multiconfiguration reference configuration interaction (MRCI) method with the Davidson correction \cite{mrci1,mrci2,mrci3} and the aug-cc-pV5Z basis set \cite{Kendall1992,Woon1993,Woon1994}. We obtain appropriate orbitals by multi-state multiconfiguration self-consistent field (MCSCF) calculations \cite{knowles1985,kreplin2019,kreplin2020,werner1985second}. The active space is composed of 8 electrons distributed over 5 orbitals in the $A'$ symmetry and 1 orbital in the $A''$ symmetry. 

The potential energy surfaces $V_\textrm{int}(R,\theta)$ are anisotropic and can be expanded in the basis of Legendre polynomials, $P_{\lambda}(\cos\theta)$, as
\begin{equation}
V_\textrm{int}(R,\theta) = \sum_{\lambda=0}^{\lambda_\textrm{max}} V_{\lambda}(R) P_{\lambda}(\cos\theta),
\end{equation}
where $V_{\lambda}(R)$ are the expansion coefficients dependent on $R$. Potential energy surfaces $V_\textrm{int}(R,\theta)$ are calculated on a two-dimensional grid of 50 points in $R$ between 5 and 35 Bohr and 15 points in angle $\theta$ between $0^{\circ}$ and $180^{\circ}$ chosen to be the roots of the Legendre polynomial of the order of 15.

\label{sec:abtheory}

At long range, the atom+molecule potential is dominated by attractive van der Waals interactions of the form
\begin{equation}
V(R,\theta)\approx-\frac{C_{6,0}}{R^{6}}-\frac{C_{6,2}}{R^6}P_{2}(\cos\theta)\,,   
\end{equation}
where other Legendre components have leading $C_{n,m}$ with $n>6$. The isotropic $C_{6,0}$ coefficient is the sum of the dispersion and induction contributions. The dispersion is an intermonomer correlation effect and corresponds to the interaction of fluctuating instantaneous dipole moments. The isotropic dispersion coefficient calculated from the Casimir-Polder formula is \citep{KORONA_2006}
\begin{equation}
C_{6,0}^\textrm{dis}=\frac{3}{\pi}\int_{0}^{\infty}\bar{\alpha}^\textrm{AlF}(i\omega) \bar{\alpha}^\textrm{He}(i\omega) d\omega\,, \label{eq:C60}
\end{equation}
where $\bar{\alpha}^{X}$
is the mean dynamic dipole polarizability of $X$. The anisotropic $C_{6,2}^\textrm{dis}$ coefficient is given by
\begin{equation}
 C_{6,2}^\textrm{dis} = \frac{1}{\pi}\int_{0}^{\infty}\Delta\alpha^\textrm{AlF}(i\omega) \bar{\alpha}^\textrm{He}(i\omega) d\omega\,,
\end{equation}
where $\Delta\alpha^\text{AlF}$ is the anisotropy of the polarizability of the AlF molecule defined as the difference between its parallel $\alpha^\text{AlF}_{\parallel}$ and the perpendicular $\alpha^\text{AlF}_{\perp}$ dynamic dipole polarizabilities:  $\Delta\alpha^{X}  = \alpha^\text{AlF}_{\parallel}-\alpha^\text{AlF}_{\perp}$. 

For AlF in the ground state, we calculate polarizabilities using the coupled cluster polarization propagator method~\cite{Jeziorki1993,Korona2011,KORONA_2006,Korona2006,Modrze2014} and obtain $C_{6,0}^\textrm{dis}=18.97\,E_{h}a_{0}^{6}$ and $C_{6,2}^\textrm{dis}=-0.13\,E_{h}a_{0}^{6}$.  For AlF in the excited states, we scale the ground-state isotropic and anisotropic dispersion coefficients by the ratio of the corresponding molecular static isotropic and anisotropic polarizabilities.

The induction effect is due to the polarization of one monomer due to the static field of the other monomer, in this case, AlF. The induction energy is determined by the permanent multipole moments and static polarizabilities of the monomers. The induction contribution to the $C_{6,m}$ coefficients can be written as
\begin{equation}
C_{6,0}^\textrm{ind} = C_{6,2}^\textrm{ind}=\mu_\textrm{AlF}^{2}\bar{\alpha}^\textrm{He}(0)
\end{equation}
where $\mu_\textrm{AlF}$ is the permanent electric dipole moment of AlF. The static polarizability of He, $\bar\alpha^\textrm{He}(0)$, is taken from Ref.~\citep{MASILI_2003}. The value of $C_{6,0(2)}^\textrm{ind}$ is calculated to be 0.48 $E_{h}a_{0}^{6}$.

All electronic structure calculations are performed with the Molpro~\cite{Molpro,MOLPRO-WIREs,molpromain} and \textsc{mrcc}~\cite{mrcc,Kallay2020,Kallay_2005,kallay2005,Kallay2008} packages of \textit{ab initio} programmes.

\subsection{Collision theory}
\label{sec:ct}

The essence of buffer gas cooling is the thermalization of AlF molecules by colliding with He atoms. Thus, scattering calculations are necessary to understand the possible outcomes of cooling AlF by He.

The total Hamiltonian of the atom+rigid-rotor system in the center-of-mass frame under the Born-Oppenheimer approximation is 
\begin{equation}
\hat{H}=-\dfrac{\hbar^{2}}{2\mu}\nabla^{2}_{R}+\dfrac{\hbar^{2}\hat{L}^{2}}{2\mu R^{2}}+V_\textrm{int}(R,\theta)+ B_{0} \hat{j}^{2}.
\end{equation}
 The first term is the component of the kinetic energy operator in the scattering coordinate $R$. The second term is the centrifugal component, where $\hat{L}$ is the rotational angular momentum of He and AlF around each other, with the quantum number $L$. $V_\textrm{int}(R,\theta)$ is the interaction potential of the colliding systems, as calculated in the previous section, and $\mu$ is the collisional reduced mass.
The states of the free molecule are eigenfunctions of the rotational Hamiltonian, $H_\textrm{rot}$ = $B_{0} \hat{j}^{2}$, with the rotational quantum number $j$ and energy $j(j+1)B_{0}$, where $B_{0}=0.55$ cm$^{-1}$ \cite{RFBarrow_1974} is the rotational constant of AlF. We neglect the small differences in rotational constant for different electronic states.

The expansion coefficients $V_\lambda(R)$ of the interaction potentials are interpolated and extrapolated using a reciprocal power reproducing kernel Hilbert space method (RP-RKHS)~\cite{Ho1996}. This produces potentials with asymptotic forms as a sum of terms with different inverse powers depending on the parameters used. Terms with different values of $\lambda$ in the expansion have different leading terms, and we have chosen parameters in the method to give the correct leading powers up to $R^{-10}$. For the long-range coefficients calculated in Sec.~\ref{sec:abtheory}, we use the method of Ref. \cite{Ho2000} to fix the extrapolation to these values.

We perform coupled-channel scattering calculations using the \textsc{molscat} package \cite{Hutsonmolscat2019, mbf-github:2020}. The angular component of the wave function is expanded in the total angular momentum basis \cite{arthursPMPES1960}, limited by $j_\textrm{max}=12$. Coupled equations are propagated using the Manolopoulos diabatic modified
log-derivative \cite{Manolo1986} and the Alexander-Manolopoulos Airy propagator \cite{Millard1987}. The solutions are matched to asymptotic boundary conditions. $S$ matrices are extracted, and cross-sections are calculated using the usual methods.

Bound states are calculated using the \textsc{bound} package \cite{mbf-github:2020,Hutsonbound2019}. This is closely related to \textsc{molscat} and uses similar coupled-channels methods, but matches to bound-state boundary conditions and solves for eigenenergies.

\section{Results and Discussion}
\label{sec:col}

\subsection{Potential energy surfaces}
\label{sec:abresults}

\begin{table}[b]
\caption{Equilibrium geometries $R_e$, $\theta_e$ and equilibrium well depths~$D_e$ of the AlF+He complex in different electronic states. Values for global (gm) and local (lm) minima are reported.}
\label{tab:2} 
\begin{ruledtabular}
\begin{tabular}{lllll}
state & minima & $R_e\,$(bohr) & $\theta_e$~(degree) & $D_e\,$(cm$^{-1}$) \\
\hline
$X^{1}A'$ & gm & 7.71 & 180 & 24.9\\
 & & 7.1 \cite{Karra2022} & 180 \cite{Karra2022} & 22 \cite{Karra2022}\\
 & & 7.75 \cite{GOTOUMAPSS2011} & 180 \cite{GOTOUMAPSS2011} & 24.056 \cite{GOTOUMAPSS2011}\\
 & lm & 10.06 & 0 & 8.0 \\
$b^{3}A'$ & gm & 7.17 & 141.4 & 27.5\\
          & lm & 8.68 & 0 & 21.6\\
          & lm & 7.59 & 180 & 27.2\\
$a^{3}A''$ & gm & 6.00 & 85.4 &54.1\\
           & lm & 7.59 & 180 & 27.2\\
$A^{1}A'$  & gm & 7.62 & 180 &24.0\\
           & lm & 9.05 & 0 & 22.1  \\
$B^{1}A''$ & gm & 7.62 & 180 &24.0\\
           & lm &9.05 & 0 & 22.1 \\
\end{tabular}
\end{ruledtabular}
\end{table}

The equilibrium geometries and corresponding well depths for the studied electronic states of AlF+He are collected in Table~\ref{tab:2}. The calculated well depths for global minima range from 21 to 28~cm$^{-1}$, except for the $a^{3}A''$ state. They are only slightly deeper than that of a comparable YbF+He complex (21.88$\,$cm$^{-1}$~\cite{Tscherbul2007}) and lie on the lower end of the typical interaction strength scale for interactions between neutral molecules and He~\cite{Borocci2020}. The $a^3A''$ state is a notable exception, with a potential well depth approximately twice as deep, placing it near the higher end of the He+neutral molecule interaction strength scale~\cite{Borocci2020}. This significant difference indicates that the interaction strength between AlF and He can be controlled by electronic excitation.

\begin{figure}
\includegraphics[width=\columnwidth]{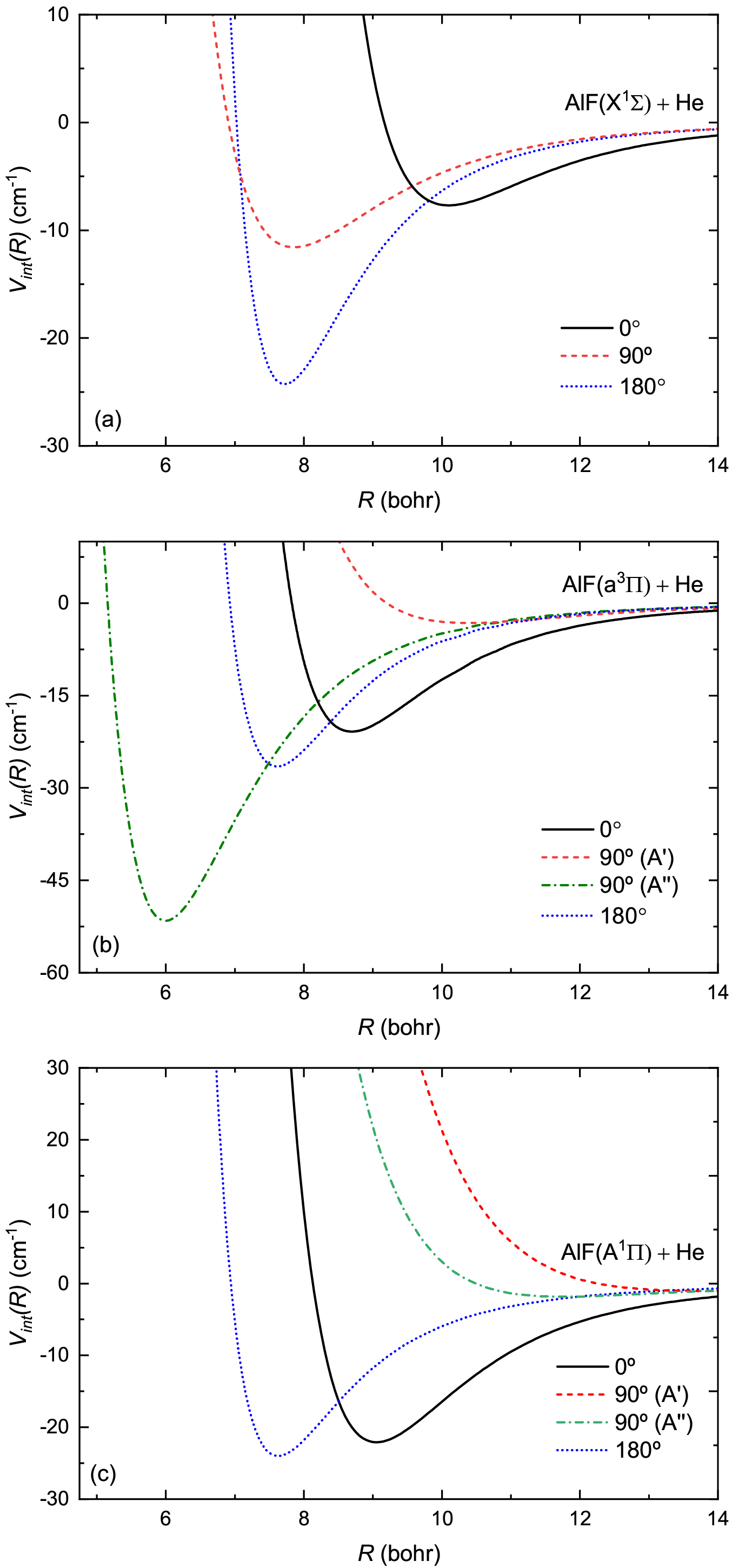}
\caption{One-dimensional cuts through the potential energy surfaces for the (a) $X^{1} A'$, (b) $a^{3}A''$ and $b^{3}A'$, and (c) $A^{1}A'$ and $B^{1}A''$ electronic states of AlF+He in linear and perpendicular orientations.}
\label{fig:2}
\end{figure}

Figure~\ref{fig:2} presents one-dimensional cuts through the potential energy surfaces for the $X^{1} A'$, $a^{3}A''$, $b^{3}A'$, $A^{1}A'$, and $B^{1}A''$ electronic states of AlF interacting with He in two linear configurations ($\theta=0^{\circ}$ and $180^{\circ}$) and one perpendicular configuration ($90^{\circ}$). For the ground state of AlF, the linear configuration with helium near fluorine displays a potential well that is twice as deep as that of both the perpendicular and alternate linear configurations, occurring at a significantly larger intermonomer separation. Electronic excitation of AlF allows helium to come closer to aluminum and reduces the differences in the well depths for two linear configurations. This illustrates how the shape of the interaction potential can be controlled by varying the electronic state of AlF.

\begin{figure*}
\centering
\includegraphics[width=2\columnwidth]{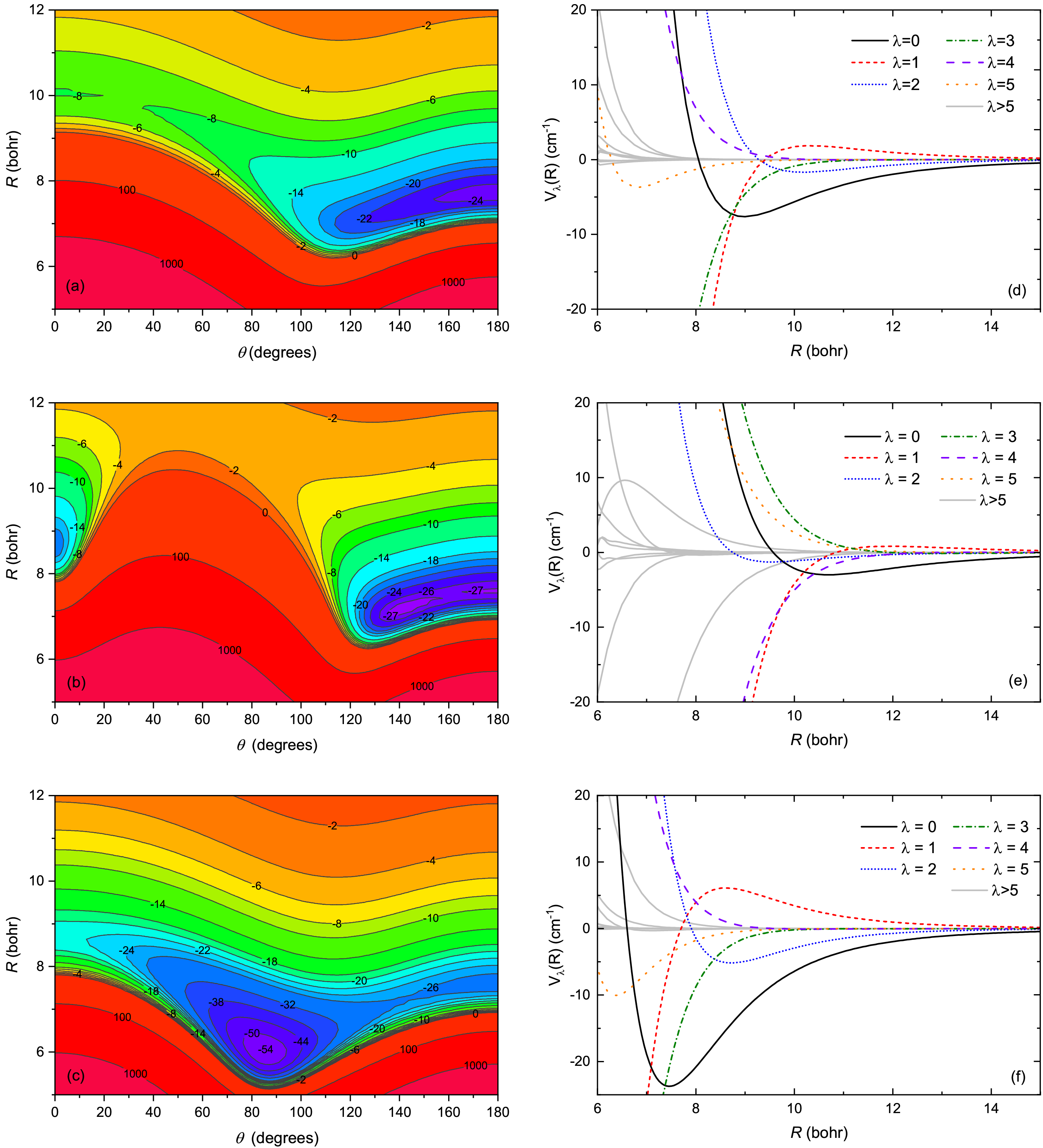}
\caption{Two-dimensional potential energy surfaces and corresponding Legendre components for the (a,d) $X^{1}A'$, (b,e) $b^{3}A'$, and (c,f) $a^{3}A''$ electronic states of AlF+He.}
\label{fig:3}
\end{figure*}

\begin{figure*}[ht]
\begin{center}
\includegraphics[width=2\columnwidth]{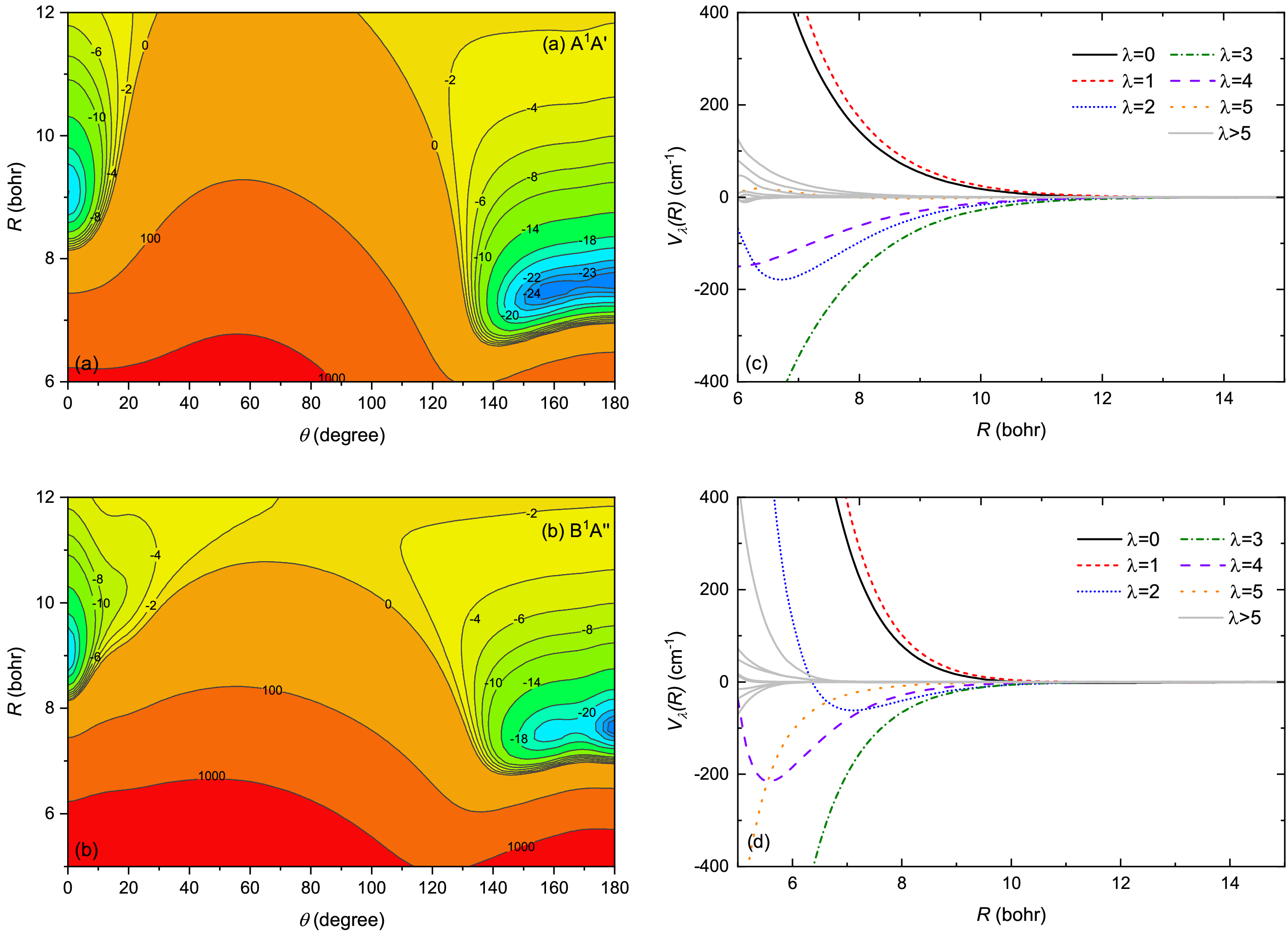}
\end{center}
\caption{Two-dimensional potential energy surfaces and corresponding  Legendre components for the (a,c) $A^{1}A'$ and (b,d) $ B^{1}A''$ electronic states of AlF+He.}
\label{fig:4}
\end{figure*}

Figures~\ref{fig:3} and~\ref{fig:4} show two-dimensional contour plots and corresponding Legendre components of the interaction potentials for the $X^{1}A'$, $a^{3}A''$, $b^{3}A'$, $A^{1}A'$, and  $B^{1}A''$ electronic states of AlF+He. The anisotropic nature of the studied interatomic interactions is clearly visible, with all surfaces showing a strong orientation dependence. The global minima of the $a^{3}A''$ and $b^{3}A'$ states occur at non-linear geometries in contrast to the singlet states that have linear equilibrium geometries. At the global minima of the $X^{1}A'$, $b^{3}A'$,  $A^{1}A'$, and  $B^{1}A''$ states, the helium atom is near the fluorine, which is the electronegative side of the molecule. A similar situation was reported for YbF+He~\cite{Tscherbul2007}.

For the $X^{1}A'$, $b^{3}A'$, $A^{1}A'$, and $B^{1}A''$ states, there is a local minimum in the linear configuration where He reaches the molecule from the side of Al. The depths of these minima are approximately 8~cm$^{-1}$, 21~cm$^{-1}$, 22.1~cm$^{-1}$ and 22.1~cm$^{-1}$ for $X^{1}A'$, $b^{3}A'$, $A^{1}A'$ and $B^{1}A''$ states, respectively. Only the $a^{3}A''$ state does not show a minimum in this orientation. The local minimum for the $X^{1}A'$ state was not reported in the previous studies, \cite{GOTOUMAPSS2011,Karra2022}  as it only emerges after including the full triple correction. This minimum is quite shallow, with the barrier between global and local minima having energy only a fraction of cm$^{-1}$ higher than the local minimum. For the $a^{3}A''$ and $b^{3}A'$ states, we see a local minimum with a depth of 27.2 cm$^{-1}$ in the linear configuration, where He reaches the molecule from the side of F, similar to the global minima of the ground states. The examination of the potential energy surfaces provides insights into the nature of interactions within the AlF+He complex.

\begin{figure}
\begin{center}
\includegraphics[width=\columnwidth]{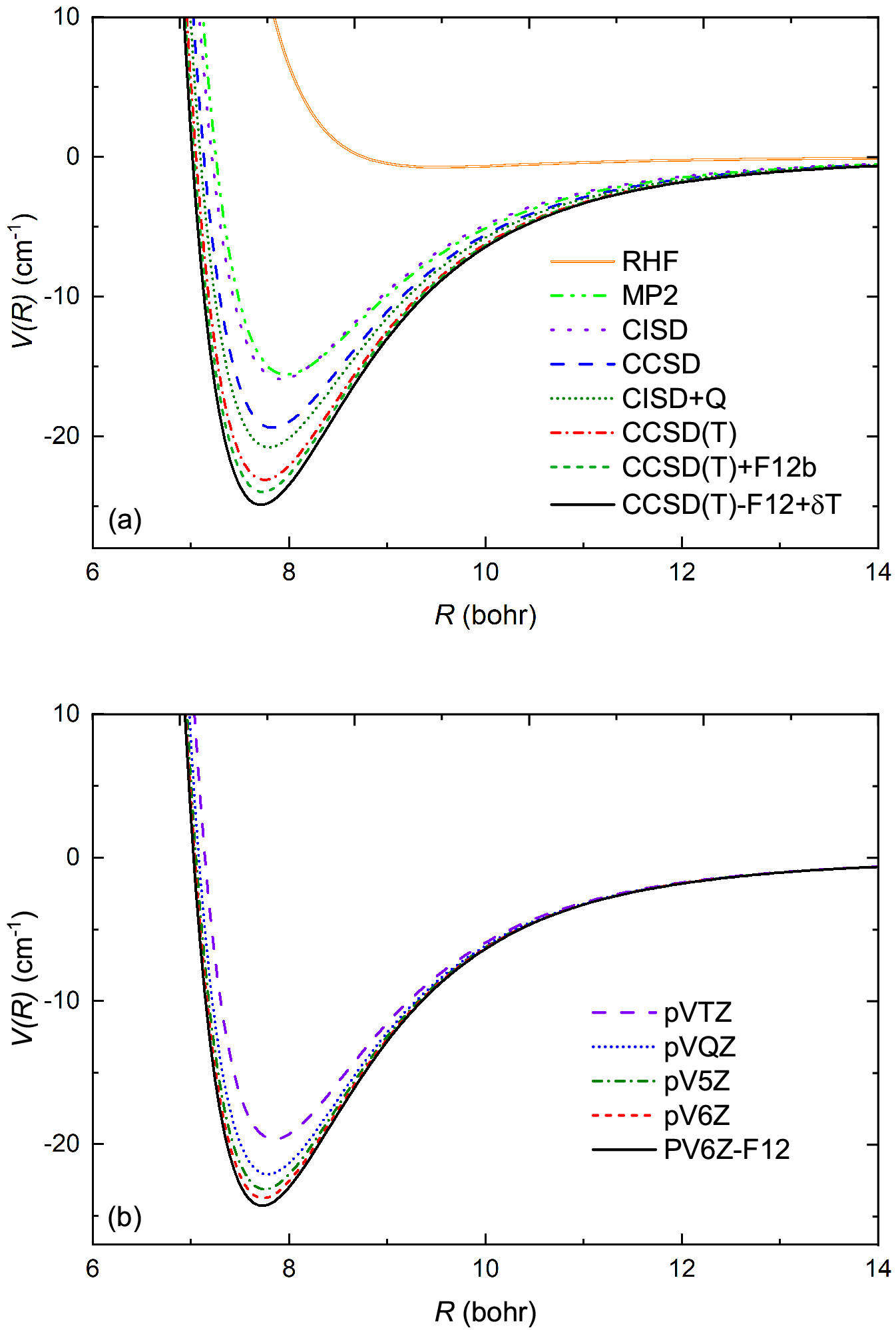}
\end{center}
\caption{The convergence of the interaction energy of AlF+He in the $X^{1}A'$ state at $\theta=180^{\circ}$ (a) for different \textit{ab initio} methods using the aug-cc-pV5Z basis set and (b) for the aug-cc-pV$X$Z basis set with increasing cardinal number $X$ using the CCSD(T) method.}
\label{fig:6}
\end{figure}

\begin{figure} 
\begin{center}
\includegraphics[width=\columnwidth]{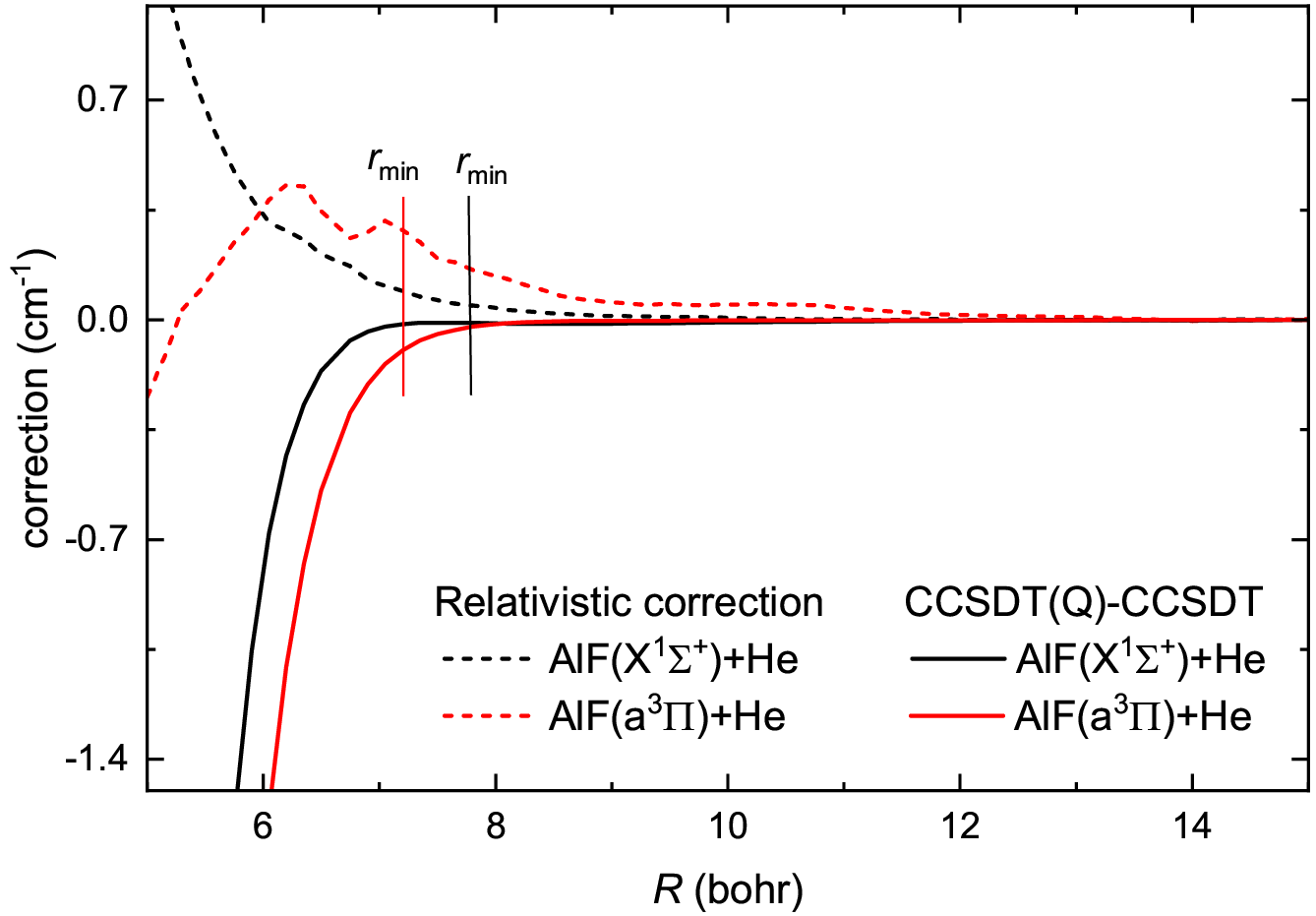}
\end{center}
\caption{The effect of including noniterative quadruple excitations within the CCSDT(Q) method and the relativistic correction within the Douglas-Kroll-Hess Hamiltonian on the interaction energy of AlF+He in the $X^{1}A'$ state at $\theta=180^{\circ}$. }
\label{fig:5}
\end{figure}

The accuracy of our electronic structure calculations may be affected by (a) an incomplete orbital basis set, (b) an inadequate description of the correlation energy, and (c) the neglect of relativistic effects~\cite{GronowskiPRA20,GebalaPRA23}. The correct description of the electron correlation is crucial for AlF+He, similar to other systems dominated by the van der Wals interaction. The convergence of interaction energy for the $X^{1}A'$ state with the quality of the wavefunction is analyzed in Fig.~\ref{fig:6}(a). The Hartree-Fock method, which is a mean-field method, produces a repulsive potential. The second-order perturbation theory, MP2, reproduces about 75\% of the correlation energy. The depth of the potential predicted at the CCSD level is 19.4~cm$^{-1}$. Further perturbative inclusion of triple excitation deepens the potential by 3.7$\,$cm$^{-1}$ while the full triple correction improves it further by 0.62~cm$^{-1}$. The perturbative treatment of the triple excitation is responsible for approximately 85\% of the total triple contribution. When calculating the entire potential energy surface, we neglect the excitations higher than triple in the coupled cluster expansion. We estimate that contribution as a difference between CCSDT(Q) and CCSDT potential, and it turns out to be relatively small, as presented in Fig.~\ref{fig:5}. It alters the interaction energy by 0.5~cm$^{-1}$ near the classical turning point and fastly decreases with the intermonomer distance to more than ten times lower value (0.045~cm$^{-1}$) at the global minimum. The reported potential energy surface also neglects the core and core-valence correlation. Including the core and core-valence correlation changes the interaction energy by 0.02~cm$^{-1}$ at the equilibrium geometry, as we estimate at the CCSD(T) level of theory.

Basis sets with increasing cardinal numbers are used to study convergence toward the complete basis limit. As shown in Fig.~\ref{fig:6}(b), the interaction strength for the $X^{1}A'$ state increases monotonically as the basis set size increases. The difference in the well depth of the potentials predicted by CCSD(T)-F12 using aug-cc-pV5Z and aug-cc-pV6Z basis sets is 0.28~cm$^{-1}$. For sextuple zeta basis sets, the difference between CCSD(T) and CCSD(T)-F12 is about 0.5~cm$^{-1}$. The complete basis set limit estimated using CCSD(T) is deeper by 0.3~cm$^{-1}$ than the depth of the well depth estimated by CCSD(T)-F12.

The system under study is relatively light, so the relativistic effects are relatively small. This can be seen in the CCSD(T)/aug-cc-pV5Z calculations with the Douglas-Kroll-Hess Hamiltonian included up to the third order presented in Fig.~\ref{fig:5}. The relativistic correction ranges from $-0.1$ to $-0.05$ cm$^{-1}$ around the global minimum and decreases with increasing intermonomer distance.

Finally, we test the accuracy of the rigid rotor approximation by optimizing the structures of AlF and AlF+He using CCSD(T)/aug-cc-pV5Z. The interaction with He decreases the equilibrium bond length of AlF by 0.008~bohr. This small relaxation of the bond length changes the interaction energies by a negligible 0.02~cm$^{-1}$.

The accuracy of our calculations is predominantly limited by the basis set incompleteness in the description of the valence electron correlation. The overall uncertainty of our potential at the global minimum is estimated to be 0.3~cm$^{-1}$, which is 1.3 \% of the well depth of the ground state potential. Our ground state interaction potential exhibits well depth similar to that in Ref.~\cite{GOTOUMAPSS2011}, but it is deeper by 2.9~cm$^{-1}$ (12\%) than that reported in Ref.~\cite{Karra2022}. This discrepancy is not surprising, as the calculations reported in Ref.~\cite{Karra2022} do not reach the complete basis set limit. The older results reported in Ref.~\cite{GOTOUMAPSS2011} used the mid-bond functions, which improve the convergence of the interaction energy calculations with the basis set size. The basis set incompleteness hampers the accuracy of our potential in the $b^{3}A'$ state by 1 cm$^{-1}$, which is 3.9\% of the well depth of the $b^{3}A'$ potential and 2\% of the well depth of the $a^{3}A''$ potential, the same for the $A^{1}A''$ and $B^{1}A'$ state (MRCI/aug-cc-pV5Z) is 0.4$\,$ cm$^{-1}$ which is 1.6 \% of the depth of the potential. We expect that the MRCI computations are affected mainly by the lack of higher excitations in the configuration interaction expansion and the size inconsistency of the method. We do not have a good estimation of those contributions. Still, we may expect that it can be even of the order of 10-20\%, similar to the difference between the CCSD and CCSDT interaction energies in the ground state.

\subsection{Collision dynamics}

\begin{figure*}
\centering
\includegraphics[width=2\columnwidth]{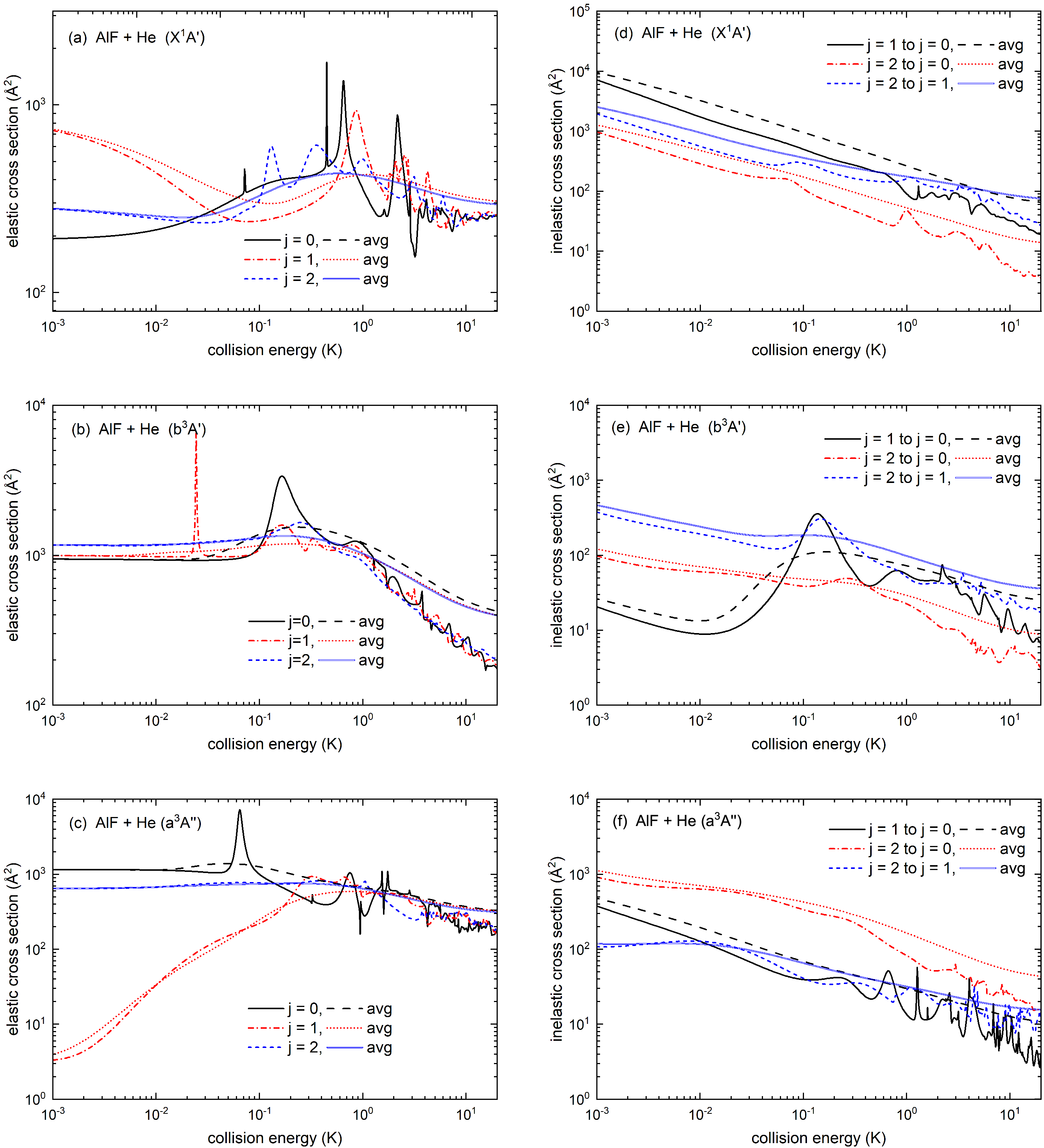}
\caption{Elastic and inelastic cross sections as a function of the collision energy for AlF+He scattering in the (a,d) $X^{1}A'$, (b,e) $b^{3}A'$, and (c,f) $a^{3}A''$ states. The cross sections of the scattering channels for the rotational states $j=0,1,2$ of AlF are shown. The corresponding curves in the same color are the thermally averaged cross sections.}
\label{fig:7}
\end{figure*}

When AlF collides with He, there are two possible outcomes: elastic collision, where there is no change in AlF's internal state, and inelastic collision, where there is a change in AlF's rotational state. Elastic collisions determine the translational thermalization of AlF, whereas inelastic collisions determine rotational thermalization.  Figure~\ref{fig:7} shows the elastic and inelastic cross sections for the AlF+He collision for the $X^{1}A'$, $a^{3}A''$ and $b^{3}A'$ states. To better understand these outcomes, we can examine the cross sections of these events.

\begin{table}[b]
\caption{Ratio of elastic to inelastic collision cross sections at 1$\,$K for the AlF+He collisions.}
\label{tab:ratio} 
\begin{ruledtabular}
\begin{tabular}{lccc}
State & $\sigma_{j=1}/\sigma_{j=1\rightarrow0}$ & $\sigma_{j=2}/\sigma_{j=2\rightarrow0}$ & $\sigma_{j=2}/\sigma_{j=2\rightarrow1}$\\
\hline
$X^{1}A'$ & 7.6 & 10.8 & 3.1\\
$ a^{3}A'' $ & 18.6 & 40.3 & 17.6\\
$ b^{3}A' $ & 36.6 & 8.9 & 25.5\\
\end{tabular}
\end{ruledtabular}
\end{table}

Let us focus on the temperatures between 100~mK to 10~K, which are the most interesting from the buffer-cell cooling perspective. A significant value of elastic cross-section is necessary to provide effective translational thermalization. In systems where inelastic cross-sections are about one to two orders of magnitude lower than elastic ones \cite{Buffer2012,Wesley2009} both translational and rotational degrees of freedom can be thermalized. The thermally-averaged elastic cross sections for the AlF+He collision (Fig.~\ref{fig:7}) lie in the range between $0.2\times 10^3$ and $2\times 10^3$~\AA$^2$, regardless of the quantum state of AlF. These calculated values are larger than the $0.1 \times 10^3$~\AA$^2$ experimentally estimated for CaH \cite{weinstein1998}, but comparable to the values calculated for that system \cite{Balakrishnan2003}. On average, buffer gas cooling can achieve a temperature range of a few hundred mK. For the $X^{1}A'$ state of AlF+He, the ratio of elastic to inelastic collision is around 7 for $\sigma_{j=1}/\sigma_{j=1\rightarrow0}$ at 1 K gradually decreasing to 0.9 at 0.2 K, around 10 for $\sigma_{j=2}/\sigma_{j=2\rightarrow0}$ at 1 K increasing to 15 near 0.5 K to 0.9 at 10 mK, around 3 for $\sigma_{j=2}/\sigma_{j=2\rightarrow1}$ at 1 K to 0.9 at 90 mK. With higher elastic collision cross sections, translational thermalization looks more prospective near 1 K. The trend is similar for other states of the AlF. The values of the ratio of elastic to inelastic collisions at 1 K are reported in Table \ref{tab:ratio}.

Let us focus on temperatures below 100~mK, which are difficult to obtain by pure buffer-cell cooling; however, we see higher sensitivity of the cross-sections on the rotational state. The elastic cross-sections are relatively constant at these temperatures, as the shape resonances gradually disappear with a decreasing temperature. 
The inelastic cross section is most significant for the state $ X^{1}A'$, resulting in fast rotational relaxation from $j=1$ to $j=0$ of AlF.
For the $X^{1}A'$ state of AlF+He, $\sigma_{j=1\rightarrow0}$ is more than an order of magnitude higher than $\sigma_{j=1}$. This means that the molecules should be cooled into the rotational ground state efficiently.
Lower values of inelastic cross-sections for triplet states can be related to lower anisotropy of the potentials, visible as a smaller difference in the shape of the potential energy curves for $\theta=0\degree$ and $\theta=180\degree$ in Fig.~\ref{fig:3}(b) and (c). Still, the inelastic and elastic cross-sections are usually in comparable range. Panel (c) of Fig.~\ref{fig:7} shows that for the $ a^{3} A'' $ state, the elastic cross section for $j=1$ is negligibly low in the low-energy limit; this is due to the scattering length being accidentally close to zero. This will inhibit translational thermalisation in this state, but the inelastic cross section is much larger, so the molecules will rapidly reach a different state where cooling could continue.

\begin{table}
\caption{Bound rovibrational levels of AlF+He in the $X^{1}A'$ state. $J$ is the total angular momentum, $p$ is the parity of the state. Energies are in cm$^{-1}$.}
\label{tab:4} 
\begin{ruledtabular}
\begin{tabular}{l|ccccc}
 $J^{p}$ &  &  &  &  &  \\
\hline
0$^{+}$ & -7.59 & -2.44 & -0.63 & & \\
1$^{+}$ & -4.77 & -0.03 & & &\\
1$^{-}$ & -7.22 & -4.84 & -2.01 & -0.81 &\\
2$^{+}$ & -6.50 & -4.13 & -1.52 & -1.16 & -0.43\\
2$^{-}$ & -3.96 & -1.45 & & &\\
3$^{+}$ & -2.75 & -0.41 & & & \\ 
3$^{-}$ & -5.43 & -3.07 & -0.71 & & \\
4$^{+}$ & -4.01 & -1.68 & & & \\
4$^{-}$ & -1.16 & & & & \\
5$^{-}$ & -2.27 & & & & \\
6$^{+}$ & -0.21 & & & & \\
\end{tabular}
\end{ruledtabular}
\end{table}

The bound rovibrational levels of the AlF+He complex in the X$^{1}A'$ state are reported in Table \ref{tab:4}. The lowest bound state supported by this potential is at -7.59 cm$^{-1}$. There are 26 bound states for the ground state in total. For the $a^{3}A''$ state with a well depth of 54 cm$^{-1}$, there are 124 bound states, which are tabulated in the Supplemental Material.

\begin{figure}
\begin{center}
\includegraphics[width=\columnwidth]{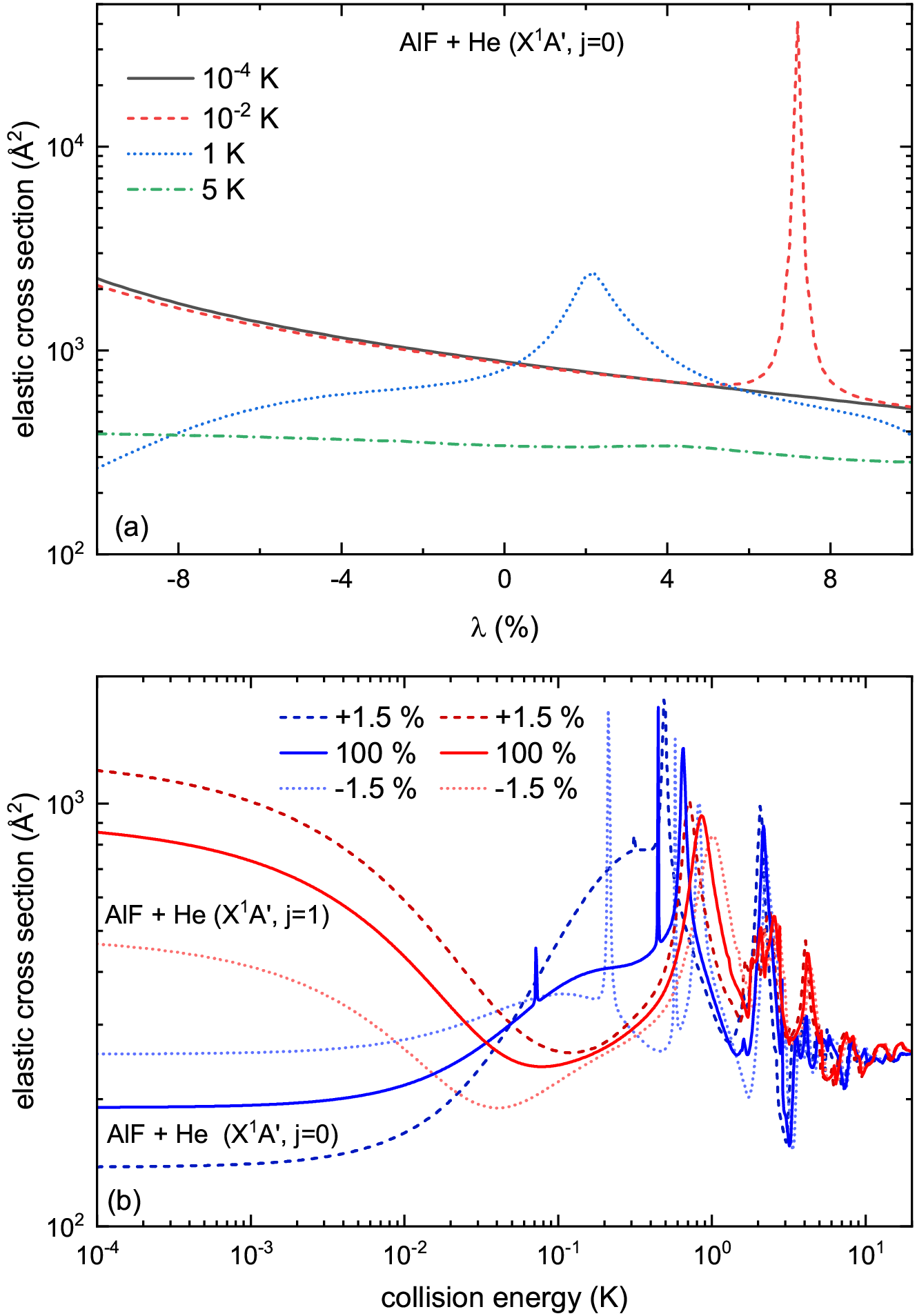}
\end{center}
\caption{(a) The sensitivity of the elastic scattering cross section to the scaling of the potential for the $X^{1}A'$ state of AlF+He for the collision energies 5 K, 1 K, 0.01 K, and 0.0001 K. (b) Elastic scattering cross-sections as a function of the collision energy for the $X^{1}A'$ potential scaled by $\pm1.5$\%.}
\label{fig:9}
\end{figure}

We study the sensitivity of our scattering results to interaction energy by scaling potential by $\pm 10\%$ and observe changes in the elastic cross-sections. Figure~\ref{fig:9}(a) shows how the cross section varies with a scaling of the potential for four energies. Here, $\lambda$ scales the potential energy as $V \rightarrow (1+\lambda)\times V$. We observe a peak in the 1 K line (Fig.~\ref{fig:9}(a)), which can be attributed to the sensitivity of the shape resonances visible around 1 K in Fig.~\ref{fig:7},(a). The uncertainty of the potential energy surface is about 2 \%, so the scattering results are not very far from the truth for all four energies. In Fig.~\ref{fig:9}(b) we show the shift in elastic cross sections for a $\pm 1.5\%$ scaling for two rotational states of AlF. For the $j=0$ rotational state of AlF, the first few shape resonances are more sensitive to the scaling of potential, but their relative sensitivity decreases for higher energies. For the $j=1$ rotational state of AlF, the curves are shifted while preserving the general shape. The scattering cross sections generally lie in the same range as predicted recently in Ref. \cite{Karra2022}, with a shift in the position of shape resonances. Although our \textit{ab initio} calculations are more accurate, the scattering results from Ref. \cite{Karra2022} are similar, which suggests that the details of the potential energy surface have a moderate effect on the scattering outcomes for this system.

\section{Conclusion}
\label{sec:summary}

We have constructed accurate potential energy surfaces using the explicitly correlated coupled-cluster and multireference configuration interaction methods for the AlF+He interactions with AlF in the ground and excited electronic states. We have found that interactions between AlF and He are weak and anisotropic. For all electronic states, we have identified global and local minima. For most of the states, the global minimum is linear and appears when He approaches AlF from the F side. The only exception from this rule is the $a^3A''$ state, however, it exhibits deep local minima for the linear AlFHe geometry, too. AlF+He is a relatively light system uninfluenced by relativistic effects. We have demonstrated that the shape and depth of the AlF+He interaction potential can be controlled by changing the electronic state of AlF. The calculated potential energy surfaces are provided in numerical form in the Supplemental Material. 

Next, we have used coupled-channel scattering theory to calculate the elastic and inelastic scattering cross sections of the AlF+He collision outcomes. The uncertainties of the potential energy surfaces have been estimated, and the sensitivity of the scattering results to the scaling of the potential has been studied to give us a better idea of the predictability of the results. The presented data can help to determine the optimal conditions for buffer gas cooling. The high ratio of elastic to inelastic collisions suggests that buffer gas cooling of AlF molecules can be efficient.

\begin{acknowledgments}
We gratefully acknowledge the Foundation for Polish Science within the First Team programme co-financed by the European Union under the European Regional Development Fund for the financial support and the Poland’s high-performance computing infrastructure PLGrid (HPC Centers: ACK Cyfronet AGH) for providing computer facilities and support within computational grant no. PLG/2023/016115. Dr. Piotr Gniewek helped us with calculating long-range coefficients and we are grateful for his support.
\end{acknowledgments}

\bibliography{AlF+He}

\end{document}